\documentclass[preprint,pra,showpacs,nofootinbib]{revtex4}
\usepackage{mathrsfs}
\usepackage{graphicx}
\usepackage{epsfig}
\usepackage{dcolumn}
\usepackage{bm}
\usepackage{amsmath,amssymb,amsthm}
\usepackage[colorlinks=true,linkcolor=blue]{hyperref}
\usepackage{subfigure}
\usepackage{booktabs}
\usepackage[mathscr]{euscript}
\usepackage{ulem}

\newcommand{\bea}{\begin{eqnarray}}
\newcommand{\eea}{\end{eqnarray}}
\newcommand{\beq}{\begin{equation}}
\newcommand{\eeq}{\end{equation}}

\def\/{\over}

\begin{document}

\title{Prohibitions caused by nonlocality \\ for Alice-Bob Boussinesq-KdV type systems}
\author{S. Y. Lou$^{1,2}$}
\affiliation{
$^{1}$\footnotesize{Center for Nonlinear Science and Department of Physics, Ningbo University, Ningbo, 315211, China}\\
$^2$\footnotesize Shanghai Key Laboratory of Trustworthy Computing, East China Normal University, Shanghai 200062, China }

\begin{abstract}
  It is found that two different celebrate models, the Korteweg de-Vrise (KdV) equation and the Boussinesq equation, are linked to a same model equation but with different nonlocalities. The model equation is called the Alice-Bob KdV (ABKdV) equation which was derived from the usual KdV equation via the so-called consistent correlated bang (CCB) companied by the shifted parity (SP) and delayed time reversal (DTR). The same model can be called as the Alice-Bob Boussinesq (ABB) system if the nonlocality is changed as only one of SP and DTR. For the ABB systems, with help of the bilinear approach and recasting the multi-soliton solutions of the usual Boussinesq equation to an equivalent novel form, the multi-soliton solutions with even numbers and the head on interactions are obtained. However, the multi-soliton solutions with odd numbers and the multi-soliton solutions with even numbers but with pursuant interactions are prohibited. For the ABKdV equation, the multi-soliton solutions exhibit many more structures because an arbitrary odd function of $x+t$ can be introduced as background waves of the usual KdV equation.\\
\leftline{}
\end{abstract}

\maketitle

\section{Introduction}
Recently, the so-called nonlocal integrable systems (or the Alice-Bob systems) have attracted many attentions of scientists because of the first paper presented by
 Ablowitz and Musslimani \cite{AM} and the possible applications for two-place physics (Alice-Bob physics) \cite{AB,ABs}. The first two-place nonlocal model is related to the places $\{x,\ t\}$ and $\{-x,\ t\}$, respectively, for the
nonlinear Schr\"odinger (NLS) equation\cite{AM}
\begin{equation}
iA_{t}+A_{xx}\pm A^2B=0,
\ \ \  B=\hat{f}A=\hat{P}\hat{C}A=A^*(-x,t), \label{Eq1}
\end{equation}
where the operators $\hat{P}$ and $\hat{C}$ are the usual parity and charge conjugation.
It is clear that the equation system \eqref{Eq1} with its conjugate form is form invariant under the discrete transformation group with the generators $\{\hat{P},\ \hat{C},\ \hat{T}\}$, where $\hat{T}$ is the time reversal operator.
In literature, the nonlocal nonlinear Schr\"odinger equation \eqref{Eq1} is also called parity-time reversal (PT) symmetric. PT symmetry plays an important role in the quantum physics \cite{2} and many other areas of physics, such as the quantum chromodynamics \cite{3}, electric circuits \cite{4}, optics \cite{5,6}, Bose-Einstein condensates \cite{7}, atmospheric and oceanic dynamics \cite{AB,Tang} etc.

It is well known that there are various correlated and/or entangled events that may happen in different times and places. To describe some types of two-place physical problems, Alice-Bob (AB) systems \cite{AB,ABs} are proposed by using the shifted parity ($\hat{P}_{s},\ \hat{P}_{s}x=-x+x_0 $), delayed time reversal ($\hat{T}_{d},\ \hat{T}_{d}t=-t+t_0$) and charge conjugate ($\hat{C}$) symmetries. If one event (A, Alice event) is correlated/entangled to another (B, Bob event), we denote the correlated relation as $B=\hat{f}A$ for suitable $\hat{f}$ operators. Usually, the event $A=A(x,\ t)$ happened at $\{x,\ t\}$ and event $B=B(x',\ t')$ happened at $\{x',\ t'\}=\hat{f}\{x,\ t\}$. In fact, $\{x',\ t'\}$ is usually far away from $\{x,\ t\}$. Hence, the intrinsic two-place models or Alice-Bob systems are nonlocal. In addition to the nonlocal nonlinear Schr\"odinger equation \eqref{Eq1}, there are many other types of two-place nonlocal models, such as the nonlocal KdV systems \cite{9,AB,ABs}, the nonlocal modified KdV systems \cite{10,11,ABs}, the discrete nonlocal NLS systems \cite{12}, the coupled nonlocal NLS systems \cite{13}, the nonlocal Davey-Stewartson systems \cite{14,15,16}, general nonlocal NLS equation \cite{TL}  and the nonlocal peakon systems \cite{ABP} including AB Xia-Qiao-Zhou (ABXQZ), AB Cammasa-Holm (ABCH), AB Degasperis-Procesi (ABDP), AB Novikov (ABN), AB FORQ (ABFORQ) and so on.

In \cite{ABs}, we proposed a series of other types of integrable AB systems including the AB-KdV systems, AB-mKdV systems, AB-KP systems, AB-sine Gordon systems, AB-NLS systems, AB-Toda systems and AB-H1 systems. Furthermore, by using the $\hat{P}_{s},\ \hat{T}_{d}$ and $\hat{C}$ symmetries, their $\hat{P}_{s},\ \hat{T}_{d}$ and $\hat{C}$ invariant muti-soliton solutions are obtained in the elegant forms for all the AB systems listed in \cite{ABs}.
 In addition, we established a most general AB-KdV equation and presented its $\hat{P}_{s},\ \hat{T}_{d}$ and $\hat{C}$ invariant Painlev\'e II reduction and soliton-cnoidal periodic wave interaction solutions \cite{AB}.

 In section II of this paper, we write down an integrable AB real system with three different nonlocal properties. One non-locality is related to the ABKdV system while the other two are associated with the ABB systems. The Lax pairs for these two kinds of AB systems are explicitly given. The multi-soliton solutions for the ABB systems are studied in section III and the possible prohibitions on multi-soliton solutions are also discussed in this section. Section IV is devoted to investigating the multi-soliton solutions of the ABKdV system. The last section is a short summary and discussion.

\section{AB integrable systems come from same equation with different non-localities}

In Ref. \cite{AB}, a special ABKdV system (after some re-scaling transformations)
\begin{eqnarray}
&&A_t=2(A+B)A_x+A_{xxx}+G(A,B),\label{ABKdVG}\\
&& B=\hat{P}_s\hat{T}_dA=A(-x+x_0,\ -t+t_0),\label{Bxt}
\end{eqnarray}
is derived from the usual KdV equation by using the so-called consistent correlated bang (CCB) \cite{CCB} with an arbitrary $\hat{P}_s\hat{T}_d$ invariant functional $G(A,\ B)=\hat{P}_s\hat{T}_dG(A,\ B)$.

In this paper, we consider a special $\hat{P}_s\hat{T}_d$ invariant selection of $G(A,\ B)$,
\begin{eqnarray}
&&G(A,\ B)=\frac12\left(1-2A-2B\right)\left(A-B\right)_x+\frac12(B-A)_{xxx}.\label{G}
\end{eqnarray}
Under the selection \eqref{G}, \eqref{ABKdVG} becomes
\begin{eqnarray}
&&A_t=\frac12(A-B+(A+B)^2+A_{xx}+B_{xx})_x. \label{ABKdV}
\end{eqnarray}
The integrability of \eqref{ABKdV} with the nonlocal condition \eqref{Bxt} is trivial because it is derived from the usual KdV equation via CCB. In fact, the ABKdV system \eqref{ABKdV} with \eqref{Bxt} possesses the following Lax pair,
\begin{eqnarray}
&&\psi_{xx}=-\frac13(U-\Lambda)\psi,\  \label{Lx}\\
&&\psi_{t}=4\psi_{xxx}+\frac16\left(7U-V\right)_x\psi
+\frac16\left(10U+2V+3E\right)\psi_x,\label{Lt}\\
&&U\equiv\left(\begin{array}{cc}
A & B \\ B & A
\end{array}\right),\ \Lambda\equiv \left(\begin{array}{cc}
\lambda & \lambda \\ \lambda & \lambda
\end{array}\right),\ V\equiv \left(\begin{array}{cc}
B & A \\ A & B \end{array}\right),\ E\equiv \left(\begin{array}{cc}
1 & -1 \\ -1 & 1 \end{array}\right). \nonumber
\end{eqnarray}
Now, an important and interesting question is, are there any other types of non-localities such that \eqref{ABKdV} is still integrable? Fortunately, we can find that if we introduce the following two new real nonlocal conditions
\begin{eqnarray}
&&B=\hat{P}_sA=A(-x+x_0,\ t), \label{Bx}
\end{eqnarray}
and/or
\begin{eqnarray}
&&B=\hat{T}_dA=A(x,\ -t+t_0),\label{Bt}
\end{eqnarray}
\eqref{ABKdV} is really integrable because of the existence of the Lax pair,
\begin{eqnarray}
&&\psi_{xxx}=-\frac12(A+B)\psi_x-\frac1{4}\left[\frac{\sqrt{-3}}3(B-A)
+A_x+B_x-4\lambda\right]\psi,\label{Lax}
\end{eqnarray}
\begin{eqnarray}
&&\psi_t=\frac{\sqrt{-3}}3\left[3\psi_{xx}+(A+B)\psi\right],\ B=\hat{P}_s^{\alpha}\hat{T}_d^{1-\alpha}A, \ \alpha=0,1. \label{Lat}
\end{eqnarray}

It should be emphasized that though the nonlocal equation \eqref{ABKdV} is same for all three non-localities \eqref{Bxt},\ \eqref{Bx} and \eqref{Bt}, their integrable properties, Lax pairs, are quite different.

From the Lax pair \eqref{Lx} and \eqref{Lt}, we know that \eqref{ABKdV} with \eqref{Bxt} is a KdV type nonlocal system. However, \eqref{ABKdV} with \eqref{Bx} or \eqref{Bt} is a Boussinesq type nonlocal equation because its Lax pair possesses the form \eqref{Lax} and \eqref{Lat}.

\section{Mult-soliton solutions for the ABB system (\ref{ABKdV}) with (\ref{Bx}) and/or (\ref{Bt})}
The multi-linear approach, especially, Hirota's bilinear method, is a powerful method to looking for multi-soliton solutions for integrable nonlinear systems. By using the standard Hirota's bilinear operator,
\begin{equation*}
D_x^mD_t^nf\cdot f \equiv \left.(\partial_x-\partial_{x'})^m(\partial_t-\partial_{t'})^nf(x,t)f(x',t')\right|_{x'=x,t'=t},
\end{equation*}
it is straightforward to prove that the AB system \eqref{ABKdV} can be changed to the following eight-linear form, \begin{eqnarray}
&&2f^2g^2(g^2D_x^4f\cdot f+f^2D_x^4g\cdot g -2g^2D_t^2f\cdot f)+3(g^2D_xD_tf\cdot f-f^2g\cdot g)^2\nonumber\\
&&\quad -3(g^2D_x^2f\cdot f-f^2D_x^2g\cdot g)^2+2f^2g^2[g^2D_x^2f\cdot f-f^2D_x^2g\cdot g -g^2D_xD_t f\cdot f\nonumber\\
&&\quad +f^2D_xD_tg\cdot g+3 (D_xD_tf\cdot f)(D_x^2g\cdot g)-3 (D_xD_tg\cdot g)(D_x^2f\cdot f)=0, \label{MLABKdV}
\end{eqnarray}
after directly substituting the transformation
\begin{equation}
A=3 (\ln f)_{xx}+3 (\ln f)_{xt},\ B=3 (\ln g)_{xx}-3 (\ln g)_{xt}, \label{A}
\end{equation}
into the ABB equation \eqref{ABKdV} and integrating once for the variable $x$ with
\begin{equation}
g=f(-x+x_0,\ t) \quad \mbox{\rm or}\quad g=f(x,\ -t+t_0). \label{g}
\end{equation}
From the relations \eqref{Bx} or \eqref{Bt}, \eqref{A} and \eqref{g}, one can readily find
that
\begin{equation}
g=f f_1(t)f_2(x+t) \label{fg}
\end{equation}
where $f_1(t)$ and $f_2(x+t)$ are arbitrary functions of $t$ and $x+t$ respectively. By using the relation \eqref{fg}, \eqref{MLABKdV} is simplified to a very simple one
\begin{equation}
(D_{t}^2-D_x^4)f\cdot f=0 \label{BLBq}
\end{equation}
which is $f_1$ and $f_2$ independent. In fact, substituting \eqref{fg} into \eqref{A}, the result is also $f_1$ and $f_2$ independent. Thus, one can directly select $f$ as the shifted parity invariant or delayed time reversal invariant function, i.e., $f_1=f_2=1$ and
\begin{equation}
f(-x+x_0,\ t)=f(x,\ t)\quad \mbox{\rm or}\quad  f(x,\ -t+t_0)=f(x,\ t). \label{even}
\end{equation}
It is clear that the only remaining thing is to solve the bilinear equation \eqref{BLBq} with the invariant condition \eqref{even} to get multi-soliton solutions.

Eq. \eqref{BLBq} is just the bilinear form of the well known Boussinesq equation. The multi-soliton solutions of the usual bilinear Boussinsq system can be written as ($f=f_n$) \cite{ABs}
\begin{eqnarray}
f_n=\sum_{\{\nu\}}K_{\{\nu\}}\cosh\left(\sum_{i=1}^n\nu_i\xi_i\right), \ \xi_i=k_ix+2\delta_ik_i^2t+\xi_{i0},\ \delta_i^2=1,\label{nsol}
\end{eqnarray}
where the summation of $\{\nu\}=\{\nu_1,\ \nu_2,\ \ldots,\ \nu_n\}$ should be done for all
permutations of $\nu_i=1,\ -1,\ i=1,\ 2,\ \ldots,\ n$, and
\begin{equation}
K_{\{\nu\}}=\prod_{i<j}a_{ij},\quad a_{ij}^2=2k_i^2+2k_j^2-k_ik_j(\delta_i\delta_j+3\nu_i\nu_j).\label{K}
\end{equation}

Generally, the solution \eqref{nsol} is not shifted parity or delayed time reversal invariant. Before to give a general result, we consider the possible constraints on \eqref{nsol} because of the condition \eqref{even} for small $n$.

For $n=1$, we have
\begin{eqnarray}
f_1=\cosh\left(k_1x+2\delta_1k_1^2t+\xi_{10}\right), \  \delta_1^2=1.\label{sol1}
\end{eqnarray}
From \eqref{sol1}, the invariant condition \eqref{even} becomes
\begin{eqnarray}
\cosh\left[k_1\left(x_0-x\right)+2\delta_1k_1^2t+\xi_{10}\right]=
\cosh\left(k_1x+2\delta_1k_1^2t+\xi_{10}\right),\label{ev1x}
\end{eqnarray}
or
\begin{eqnarray}
\cosh\left[k_1x+2\delta_1k_1^2\left(t_0-t\right)+\xi_{10}\right]=
\cosh\left(k_1x+2\delta_1k_1^2t+\xi_{10}\right),\label{ev1t}
\end{eqnarray}
It is not difficult to see that there is no possible nontrivial solution of \eqref{ev1x} or \eqref{ev1t} for arbitrary $\{x,\ t\}$ by selecting parameters $k_1,\ \xi_{10},\ x_0$ and $t_0$ .

For $n=2$, the solution \eqref{nsol} becomes,
\begin{eqnarray}
f_2&=&\delta_+\sqrt{2k_1^2+2k_2^2-k_1k_2(\delta_1\delta_2+3)}
\cosh\left[(k_1+k_2)x+2(\delta_1k_1^2+\delta_2k_2^2)t+\xi_{10}+\xi_{20}\right]\nonumber\\
&&
+\delta_-\sqrt{2k_1^2+2k_2^2-k_1k_2(\delta_1\delta_2-3)}
\cosh\left[(k_1-k_2)x+2(\delta_1k_1^2-\delta_2k_2^2)t+\xi_{10}-\xi_{20}\right],\nonumber \\
&&
\delta_1^2=\delta_2^2=\delta_+^2=\delta_-^2=1,\label{sol2}
\end{eqnarray}
From the expression \eqref{sol2}, we know that if we take the parameter conditions,
\begin{eqnarray}
k_2=\pm k_1,\ \delta_2=\mp\delta_1,\ \xi_{10} = -\frac{k_1}2(2\delta_1k_1t_0+x_0),\ \xi_{20} = -\frac{k_2}2(x_0+2\delta_2k_2t_0),\label{kd2}
\end{eqnarray}
then two-soliton solution \eqref{sol2} is restricted to the simple form
\begin{eqnarray}
f_2&\sim & \delta_{\pm}
\cosh\left[2k_1\left(x-\frac{x_0}2\right)\right]+2\delta_{\mp}
\cosh\left[4k_1^2\left(t-\frac{t_0}2\right)\right] \label{sol21}
\end{eqnarray}
up to a neglected constant factor $\sqrt{2}k_1$.

It is not difficult to check that \eqref{sol21} is the only shifted parity or delayed time reversal invariant form of \eqref{sol2}.

The conditions $k_2=\pm k_1,\ \delta_2=\mp\delta_1$ for the two-soliton solution \eqref{sol2} implies that two solitons can exist only for head-on collisions with the same velocities and wave numbers. In other words, because of the nonlocal conditions \eqref{Bx} or \eqref{Bt}, the pursuant interactions between two solitons are prohibited. The head on collisions with different wave numbers (and then velocities) are also prohibited.

Fig. 1 exhibits the head on collision expressed by \eqref{A} with \eqref{sol21} for the field $A$ under the parameter selections
\begin{equation}
\delta_{\pm}=\delta_{\mp}=k_1=1.\label{p2}
\end{equation}

\input epsf
\begin{figure}
\centering\epsfxsize=7cm\epsfysize=5cm\epsfbox{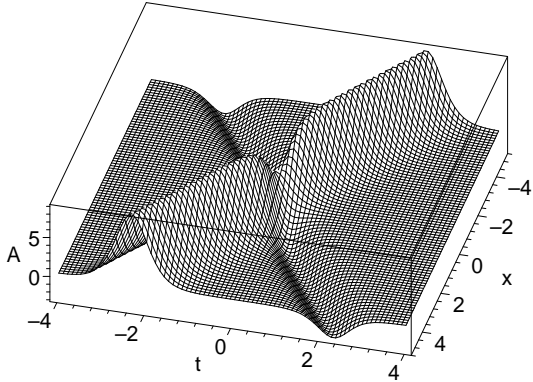}
\caption{Head on collision interaction between soliton and antisoliton for the field $A$ described by Eq. (\ref{A}) with \eqref{sol21} under the parameter
selections  (\ref{p2}).}\label{fig1}
\end{figure}

For $n=3$, the solution \eqref{nsol} possesses the form
\begin{eqnarray}
f_3&=&K_{\{\}}
\cosh\left(\xi_1+\xi_2+\xi_3\right)+K_{\{3\}}
\cosh\left(\xi_1+\xi_2-\xi_3\right)\nonumber\\
&&+K_{\{2\}}
\cosh\left(\xi_1-\xi_2+\xi_3\right)+K_{\{1\}}
\cosh\left(-\xi_1+\xi_2+\xi_3\right),\label{3sol}
\end{eqnarray}
where
\begin{eqnarray}
&&\xi_i= k_ix + 2\delta_ik_it +\xi_{i0}, \ \delta_i^2=1, \ i=1,\ 2,\ 3,\ a_{ij}^{\pm}=\sqrt{2k_i^2+2k_j^2-k_ik_j(\delta_i\delta_j\pm 3)}, \nonumber\\
&&K_{\{\}}=a_{12}^+a_{13}^+a_{23}^+, \ K_{\{3\}}=a_{12}^+a_{13}^-a_{23}^-, \ K_{\{2\}}=a_{12}^-a_{13}^+a_{23}^-, \ K_{\{1\}}=a_{12}^-a_{13}^-a_{23}^+. \label{KK}
\end{eqnarray}
In \eqref{3sol} and the following of this paper, the used notation $K_{\{i_1..i_j\}}$ means that for the permutation (the subscription of $K_{\{\nu\}}$), $\{\nu\}\equiv\{\nu_1,\ \nu_2,\ \ldots,\ \nu_n\}\equiv \{\nu_{i}=-1,\ i=i_{1},\ i_2,\ \ldots,\ i_j; \nu_{i}=1,\ i\neq i_{1},\ i_2,\ \ldots,\ i_j\} $. Thus, for $n=3$, $K_{\{\nu\}}$ in \eqref{3sol} possesses the forms $K_{\{\}}=K_{\{\nu_1=1,\nu_2=1,\nu_3=1\}},\ K_{\{1\}}=K_{\{\nu_1=-1,\ \nu_2=1,\ \nu_3=1\}}$ and so on.

After finishing some detailed analysis, one can find that there is no possible three soliton solution \eqref{3sol} because of the invariant condition \eqref{even}. Furthermore, all the soliton solutions with odd number $n$ will be prohibited by the shifted parity or the delayed time reversal nonlocality \eqref{even} or equivalently \eqref{Bx} and/or \eqref{Bt}.

For $n=4$, the solution \eqref{nsol} can be explicitly written as
\begin{eqnarray}
f_4&=&K_{\{\}}
\cosh\left(\xi_1+\xi_2+\xi_3+\xi_4\right)+K_{\{4\}}
\cosh\left(\xi_1+\xi_2+\xi_3-\xi_4\right)\nonumber\\
&&+K_{\{3\}}
\cosh\left(\xi_1+\xi_2-\xi_3+\xi_4\right)+K_{\{2\}}
\cosh\left(\xi_1-\xi_2+\xi_3+\xi_4\right)\nonumber\\
&&+K_{\{1\}}
\cosh\left(-\xi_1+\xi_2+\xi_3+\xi_4\right)+K_{\{34\}}
\cosh\left(\xi_1+\xi_2-\xi_3-\xi_4\right)\nonumber\\
&&+K_{\{24\}}
\cosh\left(\xi_1-\xi_2+\xi_3-\xi_4\right)+K_{\{23\}}
\cosh\left(\xi_1-\xi_2-\xi_3+\xi_4\right),\label{4sol}
\end{eqnarray}
where
\begin{eqnarray}
&&K_{\{\}}=a_{12}^+a_{13}^+a_{14}^+a_{23}^+a_{24}^+a_{34}^+, \ K_{\{4\}}=a_{12}^+a_{13}^+a_{14}^-a_{23}^+a_{24}^-a_{34}^-, \nonumber\\ &&K_{\{3\}}=a_{12}^+a_{13}^-a_{14}^+a_{23}^-a_{24}^+a_{34}^-, \ K_{\{2\}}=a_{12}^-a_{13}^+a_{14}^+a_{23}^-a_{24}^-a_{34}^+\nonumber\\
&&K_{\{1\}}=a_{12}^-a_{13}^-a_{14}^-a_{23}^+a_{24}^+a_{34}^+,\
K_{\{34\}}=a_{12}^+a_{13}^-a_{14}^-a_{23}^-a_{24}^-a_{34}^+\nonumber\\
&&K_{\{24\}}=a_{12}^-a_{13}^+a_{14}^-a_{23}^-a_{24}^+a_{34}^-,\
K_{\{23\}}=a_{12}^-a_{13}^-a_{14}^+a_{23}^+a_{24}^-a_{34}^-. \label{KK4}
\end{eqnarray}
After finishing some detailed analysis, for the four soliton solution \eqref{4sol}, there is only one independent selection
\begin{equation}
k_3=k_1\equiv\frac{k}2,\ k_4=k_2\equiv\frac{\kappa}2,\ \delta_3=-\delta_1=1,\ \delta_4=-\delta_2=1,\ \xi_{i0}=-\frac{k_i}2(x_0+2\delta_ik_i^2t_0). \label{kkdd}
\end{equation}
All other possible selections are equivalent to \eqref{kkdd}.

Under the restriction \eqref{kkdd}, the four soliton solution \eqref{KK4} becomes a $\hat{P}_s$ and $\hat{T}_d$ invariant form (up to a common factor $k\kappa$),
\begin{eqnarray}
&&f_4\sim 2(k^2-\kappa^2)\sqrt{(k^2+\kappa^2)^2-k^2\kappa^2}\left[\cosh\left(\eta_1\right)
+\cosh\left(\eta_2\right)+\cosh\left(\hat{T}_d\eta_1\right)
+\cosh\left(\hat{P}_s\eta_2\right)\right]\nonumber\\
&&\quad +\big(k^2+\kappa^2+k\kappa\big)(k+\kappa)^2
\cosh\left[\big(k-\kappa\big)\zeta\right] +\big(k^2+\kappa^2-k\kappa\big)(k-\kappa)^2
\cosh\left[\big(k+\kappa\big)\zeta\right] \nonumber\\
&&\quad +4\left[\big(k^2+\kappa^2\big)^2-k^2\kappa^2\right]
\cosh\left[\big(k^2-\kappa^2\big)\tau \right] +4\big(k^2-\kappa^2\big)^2
\cosh\left[\big(k^2+\kappa^2\big)\tau\right],\label{f4}\\
&&\eta_1=k\zeta
+\kappa^2\tau,\ \eta_2=\kappa \zeta
+k^2\tau,\ \zeta\equiv x-\frac{x_0}2,\ \tau\equiv t-\frac{t_0}2.
\end{eqnarray}

Fig. 2 is a plot of the head on collision interaction for the four soliton solution described by \eqref{A} with \eqref{f4} and the parameters are fixed as
\begin{equation}
x_0=t_0=0,\ k=2.5, \ \kappa=2. \label{p4}
\end{equation}

\input epsf
\begin{figure}
\centering\epsfxsize=7cm\epsfysize=5cm\epsfbox{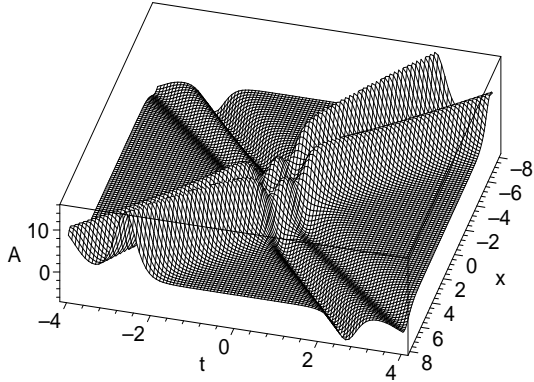}
\caption{Head on collision of two paired soliton-antisolitons exhibited in Eq. (\ref{A}) with \eqref{f4} and the parameter
selections  (\ref{p4}).}\label{fig2}
\end{figure}

It is clear that the solution \eqref{f4} is $\hat{P}_s$ and $\hat{T}_d$ invariant. The only independent selection \eqref{kkdd} for the four soliton solution implies that there are only soliton-antisoliton pairs with the same wave number but the opposite velocities.  While other types of soliton interactions such as the pursuant interaction and the head on collision with different wave numbers are prohibited by the shifted parity and or the delayed time reversal nonlocality conditions \eqref{even} or equivalently \eqref{Bx} and \eqref{Bt}. In fact, this conclusion is correct for all $n=2N$ soliton solution \eqref{nsol}.

For $n=2N$, the invariant condition of \eqref{nsol} becomes
\begin{equation}
k_{N+i}=k_i,\  \delta_{i}=-\delta_{N+i}=1,\
\xi_{i0}=-\frac{k_i}2(x_0+2\delta_ik_i^2t_0). \label{kN}
\end{equation}
Because of the paring condition \eqref{kN}, the even function property of cosh function in \eqref{nsol} and the summation is done for all possible permutations $\{\nu\}=\{\nu_1,\ \ldots,\ \nu_{2N}\}$ with $\nu_i=1,\ -1$, the solution $f_{2N}$ expressed by \eqref{nsol} with \eqref{kN} is always $\hat{O}$-invariant
\begin{equation}
\hat{O}\equiv \hat{P}_s^{\alpha}\hat{T}_d^{1-\alpha}, \ \alpha=0,\ 1. \label{O}
\end{equation}
 In other words, \eqref{nsol} with $n=2N$ and \eqref{kN} is even for both $\zeta\equiv \left(x-\frac{x_0}2\right)$ and $\tau\equiv \left(t-\frac{t_0}2\right)$.
For instance, for $N=3$ we have
\begin{eqnarray}
f_6&=&K_{\{23\}}\left[\cosh (y_{1+}) +\cosh \big(\hat{O}y_{1+}\big) \right]
+K_{\{35\}}\left[\cosh (y_{1-}) +\cosh \big(\hat{O}y_{1-}\big) \right]\nonumber\\
&&+ K_{\{13\}}\left[\cosh (y_{2+}) +\cosh \big(\hat{O}y_{2+}\big) \right]
+K_{\{34\}}\left[\cosh (y_{2-}) +\cosh \big(\hat{O}y_{2-}\big) \right]\nonumber\\
&&+K_{\{12\}}\left[\cosh (y_{3+}) +\cosh \big(\hat{O}y_{3+}\big) \right]
+K_{\{24\}}\left[\cosh (y_{3-}) +\cosh \big(\hat{O}y_{3-}\big) \right]\nonumber\\
&&+K_{\{3\}}\left[\cosh (z_{3+}) +\cosh \big(\hat{O}z_{3+}\big) \right]
+K_{\{146\}}\left[\cosh (z_{3-}) +\cosh \big(\hat{O}z_{3-}\big) \right]\nonumber\\
&&+K_{\{2\}}\left[\cosh (z_{2+}) +\cosh \big(\hat{O}z_{2+}\big) \right]
+K_{\{356\}}\left[\cosh (z_{2-}) +\cosh \big(\hat{O}z_{2-}\big) \right]\nonumber\\
&&+K_{\{1\}}\left[\cosh (z_{1+}) +\cosh \big(\hat{O}z_{1+}\big) \right]
+K_{\{346\}}\left[\cosh (z_{1-}) +\cosh \big(\hat{O}z_{1-}\big) \right]\nonumber\\
&&+K_{\{\}}\cosh\left[2(k_1+k_2+k_3)\zeta\right]
+K_{\{14\}}\cosh\left[2(k_1-k_2-k_3)\zeta\right]\nonumber\\
&&+K_{\{25\}}\cosh\left[2(k_2-k_1-k_3)\zeta\right]
+K_{\{36\}}\cosh\left[2(k_3-k_1-k_2)\zeta\right]\nonumber\\
&&+K_{\{456\}}\cosh\left[4(k_1^2+k_2^2+k_3^2)\tau\right]
+K_{\{156\}}\cosh\left[4(k_1^2-k_2^2-k_3^2)\tau\right]\nonumber\\
&&+K_{\{246\}}\cosh\left[4(k_2^2-k_1^2-k_3^2)\tau\right]
+K_{\{126\}}\cosh\left[4(k_3^2-k_1^2-k_2^2)\tau\right], \label{f6}
\end{eqnarray}
where
\begin{eqnarray}
y_{i\pm}&=&2k_i\zeta +4(k_{i+1}^2\pm k_{i+2}^2)\tau,\ i=1,2,3,\ \mod 3,\nonumber\\
z_{i\pm}&=&4k_i^2\tau+2(k_{i+1}\pm k_{i+2})\zeta, \ i=1,2,3,\ \mod 3. \label{yz}
\end{eqnarray}
Fig. 3 displays the three paired soliton-antisolitons head on collision interaction expressed by \eqref{A} with \eqref{f6} and parameter selections
\begin{equation}
x_0=t_0=0,\ k_1=-1.2, \ k_2=1.1,\ k_3=-1. \label{p6}
\end{equation}

\input epsf
\begin{figure}
\centering\epsfxsize=7cm\epsfysize=5cm\epsfbox{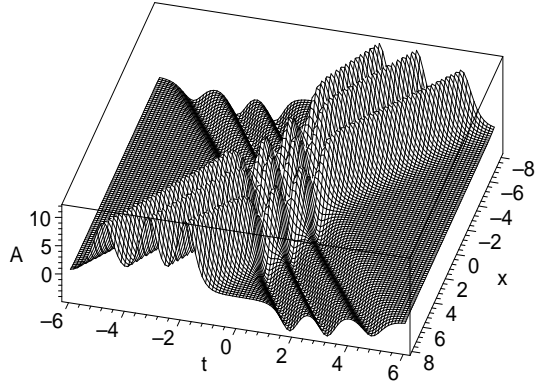}
\caption{Plots of the head on collision interactions of six solitons expressed by Eq.  (\ref{A}) with the parameter
selections  (\ref{p6}).} \label{fig3}
\end{figure}

\section{Multi-soliton solutions for the ABKdV system (\ref{ABKdV}) with (\ref{Bxt})}
For the ABKdV system \eqref{ABKdV} with the nonlocal condition \eqref{Bxt}, we can make the bilinear transformation
\begin{equation}
A=3(\ln f)_{xx}+3(\ln g)_{xt},\ B=3(\ln f)_{xx}-3 (\ln g)_{xt}. \label{BLT}
\end{equation}
Substituting the transformation relation \eqref{BLT} into the ABKdV system \eqref{ABKdV} and its adjoint form (obtained by applying the operator $\hat{P}_s\hat{T}_d$ on \eqref{ABKdV}), we can find two separated bilinear equations
\begin{eqnarray}
&&(D_xD_t-D_x^4)f\cdot f=0,\label{BLf}\\
&&(D_xD_t-D_t^2)g\cdot g=0. \label{BLg}
\end{eqnarray}
It is interesting that the $f$ equation is just the bilinear form of the usual KdV equation while the general solution of \eqref{BLg} can be written as
\begin{eqnarray}
g=g_0(x)g_1(x+t)\sim g_1(x+t). \label{rg}
\end{eqnarray}
From the expression \eqref{BLT}, we know that $g_0(x)$ can be simply taken as $1$ without loss of generality.

Applying the nonlocal condition \eqref{Bxt}, we have
\begin{eqnarray}
\hat{P}_s\hat{T}_df=f(-x+x_0,-t+t_0)=
\frac{f\exp\left[f_1(t)x+f_0(t)\right]}{g_1(-x-t+x_0+t_0)g_1(x+t)} ,\label{rf}
\end{eqnarray}
where $f_0,\ f_1$ and $g$ are arbitrary functions of the indicated variables. Especially, $f$ may be $\hat{P}_s\hat{T}_d$-invariant if we take the following special selections
\begin{equation}
f_1(t)=f_0(t)=0,\ g_1(-x-t+x_0+t_0)=g_1(x+t)^{-1}. \label{fg1}
\end{equation}

In general, the solutions of \eqref{BLf} are not $\hat{P}_s\hat{T}_d$-invariant. Fortunately, because the expression \eqref{BLT} is invariant under the transformation
\begin{equation}
f\longrightarrow \exp(Kx+\Omega t+X)f \label{ff}
\end{equation}
with arbitrary constants $K,\ \Omega$ and $X$, the multi-soliton solutions of the ABKdV equation can be written as
\begin{equation}
A=3(\ln F_N)_{xx}+3(\ln G)_{xt}, \label{NS}
\end{equation}
where $G=G(x+t)$ is an arbitrary function with the condition
\begin{equation}
\hat{P}_s\hat{T}_dG=G^{-1}, \label{CG}
\end{equation}
and
\begin{equation}
F_N=\sum_{\{\nu\}}K_{\{\nu\}}\cosh\left(\sum_{i=1}^N\nu_i\xi_i\right),\ \xi_i=k_i\left(x-\frac{x_0}2\right)+4k_i^3\left(t-\frac{t_0}2\right) \label{FN}
\end{equation}
The summation of $\{\nu\}=\{\nu_1,\ \nu_2,\ \ldots,\ \nu_N\}$ in \eqref{FN} should be done for all permutations of $\nu_i=1,\ -1,\ i=1,\ 2,\ \ldots,\ N$, and
\begin{equation}
K_{\{\nu\}}=\prod_{i<j}(k_i-\nu_i\nu_j k_j).\label{K}
\end{equation}
It is clear that \eqref{FN} is really $\hat{P}_s\hat{T}_d$-invariant.

Different from the multi-soliton solutions of the ABB system discussed in the last section for the non-localities of \eqref{Bx} and \eqref{Bt}, there is no other prohibitions for the ABKdV system with the non-locality \eqref{Bxt} except for the locations of $N$ solitons.

Fig. 4 displays one soliton solution with a periodic background wave expressed by \eqref{NS} with \eqref{FN} for $N=k_1=1,x_0=t_0=0$ and $G(x+t)$ is taken as
\begin{equation}
G(x+t)=\exp(-0.2\sin(x+t)). \label{G1}
\end{equation}

\input epsf
\begin{figure}
\centering\epsfxsize=7cm\epsfysize=5cm\epsfbox{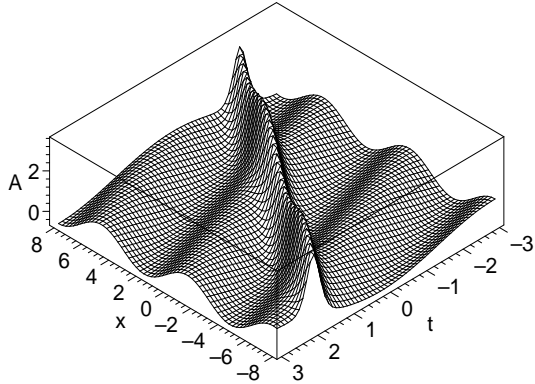}
\caption{One soliton solution \eqref{NS} with \eqref{FN} for $N=k_1=1,x_0=t_0=0$ and the periodic background wave (\ref{G1}).} \label{fig4}
\end{figure}

Fig. 5 displays the interaction between the KdV soliton and the background soliton with few peaks expressed by \eqref{NS}, \eqref{FN} for $N=k_1=1,\ x_0=t_0=0$ and
\begin{equation}
G(x+t)=\exp\left[-\int\int \mbox{\rm sech}(x+t)\sin(2x+2t)\mbox{\rm dxdt}\right]. \label{G2}
\end{equation}

\input epsf
\begin{figure}
\centering\epsfxsize=7cm\epsfysize=5cm\epsfbox{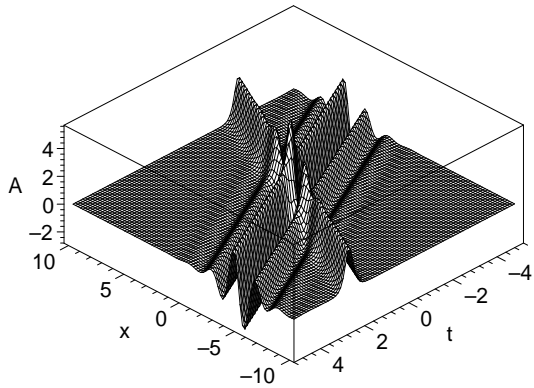}
\caption{Plot of the interaction between the usual KdV soliton and the background few cycle soliton expressed by Eq. (\ref{NS}), \eqref{FN} for $N=k_1=1,\ x_0=t_0=0$ and (\ref{G2}).} \label{fig3}
\end{figure}

The main difference between the usual KdV equation (\eqref{ABKdV} with $B=A$) and the ABKdV \eqref{ABKdV} with $B=\hat{P}_s\hat{T}_dA$) is that there is an arbitrary odd background wave
\begin{equation}
W(x+t)\equiv 3(\ln G(x+t))_{xt}=-\hat{P}_s\hat{T}_dW(x+t). \label{W}
\end{equation}

\section{summary and discussion}
It is shown that there are some really different phenomena in Alice-Bob systems (or namely nonlocal systems). From the result of this paper we can conclude that the following eight systems
\begin{eqnarray}
&&A_t=\frac12(A-B+(A+B)^2+A_{xx}+B_{xx})_x, \label{ABKdVA}\\
&&B=g_iA,\quad i=0,\ 1,\ 2,\ \ldots,\ 8, \label{ABKdVB}
\end{eqnarray}
where $g_i$ belongs to the eight order Abel group ${\cal G}$
\begin{eqnarray}
&&{\cal G}=\{g_0,\ g_1,\ \ldots,\ g_8\}\equiv \{I,\ \hat{P}_s,\ \hat{T}_d,\ \hat{C},\  \hat{P}_s\hat{T}_d,\ \hat{P}_s\hat{C},\  \hat{T}_d\hat{C},\ \hat{P}_s\hat{T}_d\hat{C}\},    \label{Group}
\end{eqnarray}
where $I=g_i^2$ for all $i$ is the identity of the group ${\cal G}$.  For $B=A$ and $B=\hat{C}A$, the models are just the real and complex local KdV equations. For $B=\hat{P}_s\hat{T}_d A$ and $B=\hat{P}_s\hat{T}_d\hat{C}A$, the models are related to the real and complex nonlocal KdV systems.
For $B=\hat{P}_sA$, $B=\hat{T}_dA$, $B=\hat{P}_s\hat{C}A$ and $B=\hat{T}_d\hat{C}A$, the models are related to the real and complex nonlocal Boussinesq systems.
For the (local and nonlocal, real and complex) KdV type equations, their integrability is guaranteed by the Lax pair \eqref{Lx} and \eqref{Lt}. For the (local and nonlocal, real and complex) Boussinesq type systems, their Lax pairs are the same as that given in \eqref{Lax} and \eqref{Lat}.

It is well known that in quantum physics, the existence of symmetries for a quantum system will lead to some prohibitions. It is interesting that the similar situation may also be found in classical physics. For the ABB systems, the model \eqref{ABKdVA} with $g_i \in \{\hat{P}_s,\ \hat{T}_d,\ \hat{P}_s\hat{C},\  \hat{T}_d\hat{C}\}$, because of the introduces  of the nonlocalities, some prohibitions are discovered. The number of solitons must be even. The odd number of solitons are prohibited. For the even number of solitons, all the solitons must be paired and every paired solitons must possess the same wave numbers, velocities but opposite moving directions. Only the head on interactions are allowed, which means the pursuant interactions are prohibited.

For the ABKdV cases, $g_i \in \{\hat{P}_s\hat{T}_d,\ \hat{P}_s\hat{T}_d\hat{C}\}$, different from the usual KdV case, for every known KdV solution, an arbitrary additional background wave described by an odd function of $x+t$ is allowed. For instance, the periodic wave and the few cycle soliton can be included as shown in Figs. 4 and 5.

Similar phenomena may be found for other kind of nonlinear nonlocal systems. For instance,
we may have a conjecture ($u\equiv A+B,\ v\equiv A-B,\ {\mbox i}=\sqrt{-1}$),
the following sixteen local and nonlocal models given by
\begin{eqnarray}
&&A_t+\frac12\left(2{\mbox i}uv-2u^3+u_{xx}-v\right)_x+2{\mbox i}uv_x=0, \label{mbq}\\
&&B=g_jA,\quad j=0,\ 1,\ 2,\ \ldots,\ 16,
\end{eqnarray}
are integrable,
where $g_j$ belongs to the sixteen order Abel group ${\cal G}$ with four generators, the shifted parity $\hat{P}_s$, the delayed time reversal $\hat{T}_d$, the charge conjugate $\hat{C}$ and the field reversal $\hat{F}$ defined by $\hat{F}A=-A$. The model \eqref{mbq} includes some interesting local and nonlocal systems related to the modified KdV and modified Boussinesq systems. The more about \eqref{mbq} and other types of possible nonlocal systems will be discussed elsewhere.

\section*{Acknowledgements}
The author is grateful  to thank Professors X. Y. Tang, D. J. Zhang, Z. N. Zhu, Q. P. Liu, X. B. Hu, Y. Q. Li and Y. Chen for their helpful discussions. The work was sponsored by the Global Change Research
Program of China (No.2015CB953904),  Shanghai Knowledge Service Platform for Trustworthy Internet of Things (No. ZF1213), the National Natural Science Foundations of China (No. 11435005) and K. C. Wong Magna Fund in Ningbo University.

\end{document}